\documentclass[fleqn,a4paper,12pt]{article}
\usepackage{amssymb}
\usepackage{amsmath}
\usepackage[dvips]{graphicx}
\usepackage{placeins}
\usepackage{lscape}
\usepackage{a4}
\usepackage{epsfig}

\renewcommand{\bar}{\overline} 
\newcommand{\bab}{\end{gather}}
\newcommand{\ri}{{\mathrm i}}
\newcommand{\p}{\partial}
\newcommand{\bea}{\begin{array}}
\newcommand{\eea}{\end{array}}
\newcommand{\beg}{\begin{gather}}

\catcode`@=11 \long
\def\@caption#1[#2]#3{\par\addcontentsline{\csname
ext@#1\endcsname}{#1} {\protect\numberline{\csname
the#1\endcsname}{\ignorespaces #2}} \begingroup \small
\@parboxrestore \@makecaption{\csname fnum@#1\endcsname}
{\ignorespaces #3}\par \endgroup} \catcode`@=12

\renewcommand{\bar}{\overline}

\newcommand{\la}{\label}

\catcode`@=11 \long
\def\@caption#1[#2]#3{\par\addcontentsline{\csname
ext@#1\endcsname}{#1} {\protect\numberline{\csname
the#1\endcsname}{\ignorespaces #2}} \begingroup \small
\@parboxrestore \@makecaption{\csname fnum@#1\endcsname}
{\ignorespaces #3}\par \endgroup} \catcode`@=12

\begin{document}

\allowdisplaybreaks
 \begin{titlepage} \vskip 2cm

\begin{center} {\Large\bf Superintegrable and shape invariant systems with position dependent mass}

 \vskip 3cm {\bf {A. G. Nikitin }\footnote{E-mail:
{\tt nikitin@imath.kiev.ua} }
\vskip 5pt {\sl Institute of Mathematics, National Academy of
Sciences of Ukraine,\\ 3 Tereshchenkivs'ka Street, Kyiv-4, Ukraine,
01601\\}}\end{center}
\vskip .5cm \rm
\begin{abstract} {Second order integrals of motion for 3d quantum mechanical
systems with position dependent masses (PDM) are classified. Namely,
all PDM  systems are specified which, in addition to their rotation
invariance, admit at least one second order integral of motion. All
such systems appear to be also shape invariant and exactly solvable.
Moreover, some of them possess the property of double shape
invariance and can be solved using two different superpotentials.  Among
them there are systems with  double shape invariance which present
nice bridges between the Coulomb and isotropic oscillator systems.

A simple algorithm for calculation the discrete spectrum and the
corresponding state vectors for the considered PDM systems is
presented and applied  to solve five of the found systems.}
\end{abstract}
\end{titlepage}
\section{Introduction\label{int}}
There are three differently defined global properties which are possessed by some quantum mechanical systems: superintegrability, supersymmetry and exact solvability.

A system with $d$ degrees of freedom is called superintegrable if it admits more than $d$ integrals of motion (including Hamiltonian). Moreover, exactly $d$ of them should commute each other. The maximal admissible number of integrals of motion is equal to $2d-1$, and the related systems are called maximally superintegrable.

The system is treated as supersymmetric in two cases: when its integrals of motion form a superalgebra, or (and) the Hamiltonian has a specific symmetry with respect to the Darboux transform,   called shape invariance.

Maximally superintegrable or shape invariant systems as a rule are exactly solvable. It means that all their energy levels can be
calculated algebraically, and the corresponding wave functions can be found in explicit form.

On the other hand, there exist a tight coupling between the maximal superintegrability and shape invariance \cite{Lym}. The list of systems which are both supersymmetric and shape invariant includes the Hydrogen atom, isotropic harmonic oscillator, 2d and 3d superintegrable systems with spin \cite{Mart}, \cite{N5}, \cite{N1}, arbitrary dimensional systems with spin 1/2 \cite{N2}, and others. New examples of such coupling are presented in the current paper.

There are various reasons to search for superintegrable and shape invariant systems. First, many of such systems are very important from physical viewpoint. Secondly, as a rule they can be solved analytically in a way free of uncertainties generated by various approximate approaches.     In addition,  these systems present a  nice field for application of symmetries in physics.

The systematic search for integrable and superintegrable systems in quantum mechanics started with fundamental papers \cite{wint1} and \cite{BM} where the second order integrals of motion for planar quantum mechanical systems have been classified. We will not discuss a very inspiring history of this research which is still continuing now, see survey \cite{wint01}. Let us only mention that there is a great  number of papers devoted to this subject. In particular, the contemporary  field of superintegrable models includes the systems with  spin \cite{w6}, \cite{w7}, \cite{w8},  \cite{N5}, \cite{N1},\cite{N6}. For the classification of shape invariant systems with spin see \cite{N3} and \cite{N4}.

In the present paper we continue the  search for  superintegrable
Schr\"odinger equations with position dependent mass (PDM), started
in paper \cite{NZ}. In contrast with the systems with a constant
mass, superintegrability aspects of such equations were not studied
systematically, although special classes of maximally
superintegrable systems are well known, see, e.g., \cite{Bala2}--
\cite{Rag3} and references cited therein. Let us remind that  just
the PDM systems are intensively used in modern physics. They are
applied for modeling of condensed-matter systems, namely,
semiconductors \cite{Roz}, \cite{1}, quantum liquids \cite{7} and
metal clusters \cite{13}, quantum wells, wires and dots \cite{2},
\cite{3} and many, many others. In addition, thanks to the
additional randomness connected with a non-fixed mass, the PDM
systems present a much more reach field for application of symmetry
methods than the systems with constant masses.

The present paper includes the complete classification of 3d PDM
systems which are invariant with respect to the rotation group and
admit second order integrals of motion. Up to equivalence,
twenty  such systems are specified. Moreover, sixteen of them  are defined up to
an arbitrary parameter, and four of them include pairs of arbitrary
parameters.

 We will show that  all obtained  systems are  both maximally
 superintegrable, shape invariant and exactly solvable. In other words, we
 prove that if a PDM system is rotationally invariant and admits second
 order integrals of motion additional to polynomials in angular momentum and dilatation,
 it possesses all global properties discussed at the beginning
 of this paper. Using these properties, we calculate the energy
 spectra for some of the obtained systems  and construct their
 solutions explicitly. We also present a simple algorithm for
 construction of discrete spectrum solutions for any of the
 presented systems.

Shape invariance is a fine symmetry which presents very convenient tools
for constructing exact solutions of systems which possess this property.
We will see that the
PDM systems include ones which have doubled hidden supersymmetry. In
other words, the corresponding radial equations are shape invariant with
respect to two different Darboux transforms.

It is necessary to note that an important and in some sense
completed class of rotationally invariant superintegrable systems
was presented in papers \cite{Bala2}--\cite{Rag3}. These systems are
quantized versions of classical
 ones, obtained starting with the Coulomb and oscillator systems in curved
 spaces. Being superintegrable and rotationally invariant, they naturally
 appear in results of our research. In particular, we specify such of them
 which admit second order integrals of motion and are shape invariant. We
 also show that there exist such rotationally invariant and superintegrable
 systems which do not belong to the class introduced in
 \cite{Bala2}--\cite{Rag3}. A more detailed discussion of these points is
 presented in section 7.
\section{Rotationally invariant PDM Schr\"odinger equations}
We will study stationary  Schr\"odinger equations with position dependent mass, which can be represented in the following form:
\begin{gather}\la{se}
   \hat H \psi=E \psi,
\end{gather}
where
\begin{gather}\la{H}\hat H=p_af({\bf x})p_a+\tilde V({\bf x}).\end{gather}
Here ${\bf x}=(x^1,x^2,x^3),$ $p_a=-i\p_a$, $V({\bf x})$ and $f({\bf x})=\frac1{2m({\bf x})}$ are arbitrary functions associated with the effective potential and inverse effective PDM, and summation from 1 to 3 is imposed over the repeating index $a$.

In paper \cite{NZ} all equations (\ref{se}) admitting at least one first order integral of motion has been classified. In particular all rotationally invariant systems with different symmetries were presented there. In general such systems are  characterized by the following
$\bf x$-dependence of $f$ and $V$:
\begin{gather}\la{fV} f=f(x),\quad \tilde V=\tilde V(x),\quad x=\sqrt{x_1^2+x_2^2+x_3^2}.\end{gather}

Hamiltonians (\ref{H}), (\ref{fV}) by construction are invariant with respect to group SO(3) whose generators are components of the angular momentum vector:
\begin{gather}\la{J}{ J_a}=\varepsilon_{abc}x_bp_c\end{gather}
where $\varepsilon_{abc}$ is  Levi-Civita tensor.

In accordance with \cite{NZ} there are exactly four such Hamiltonians which have a more extended symmetry. They are specified by the following inverse masses and  potentials:
\begin{gather}\la{fV1}f=x^2, \quad \tilde V=0,\\f=(1+x^2)^2,
\quad \tilde V=-6x^2, \la{fV2} \\f=(1-x^2)^2, \quad
\tilde V=-6x^2,\la{fV3} \\
f=x^4, \quad  \tilde V=-6x^2.
\la{fV4}\end{gather}

Hamiltonians (\ref{H}) whose arbitrary elements are fixed by equations (\ref{fV1}), (\ref{fV2}), (\ref{fV3}) and (\ref{fV4}) admit additional integrals of motion, i.e.,
\begin{gather}\la{I1}D={\bf x}\cdot{\bf p}-\frac{3\ri}2,\\ \la{I2}
N^-_{a}=\frac12(K^a-p_a),\\
\la{I3}N^+_{a}=\frac12(K^a+p_a)\end{gather}
and
\begin{gather}\la{I4}K^{a}=x^2 p^a -2x^aD\end{gather}
correspondingly.

In the case (\ref{fV1}) the rotation symmetry is extended by the scaling transformations while for cases (\ref{fV2})  and (\ref{fV3}) the corresponding integrals of motion (\ref{I2}), (\ref{J}) and  (\ref{I3}), (\ref{J}) form basises of algebras so(4) and so(1,3) correspondingly \cite{NZ}. Thanks to their high symmetry equations (\ref{se}) with potentials (\ref{fV1})--(\ref{fV3}) are exactly solvable, for their explicit solutions see \cite{NZ}.

\section{Determining equations}
Let us search for second order integrals of motion for equation (\ref{se}), i.e., for commuting with $H$ differential operators of second of the following generic form:
\begin{equation}\label{Q}
    Q=\mu^{ab}\p_a\p_b+\xi^a\p_a+\eta
\end{equation}
where $\mu^{ab}=\mu^{ba}$, $\xi^a$ and $\eta$ are functions  of $\bf
x$ and summation from 1 to 3 is imposed over all repeating indices.

By definition, operators  $Q$ should commute with $\hat H$:
\begin{equation}\label{HQ}[ \hat H,Q]\equiv  \hat H Q-Q \hat H=0.\end{equation}
Calculating the commutator and equating the coefficients for different differential operators we come to the following system of determining equations:
\begin{gather}\begin{split}\la{mmmu0}&5\left(\mu^{ab}_c+\mu^{ac}_b+ \mu^{bc}_a\right)=
\delta^{ab}\left(\mu^{nn}_c+2\mu^{cn}_n\right)+
\delta^{bc}\left(\mu^{nn}_a+2\mu^{an}_n\right)+\delta^{ac}
\left(\mu^{nn}_b+2\mu^{bn}_n\right),\end{split}\\
\la{mmmu1}
{\cal F}^a\equiv \left(\mu^{nn}_a+2\mu^{na}_n\right)f-
5\mu^{an}x_n\frac{f'}x=0,\\
\begin{split}&{\cal F}^{ab}\equiv
\left(\mu^{ab}_{nn}+\xi^a_b+\xi^b_a\right)f+
\left(\mu^{ab}_nx^n- 2\mu^{ab} -\delta^{ab}
\left(\mu^{nn}+ \xi^nx^n\right)\right)\frac{f'}x\\
&+ \left(\mu^{an}x^b+\mu^{bn}x^a+\delta^{ab}\mu^{kn}x^k\right)x^n\frac1{x^2} \left(\frac{f'}x-f''\right)=0,\end{split}
\la{mmmu2}\\\begin{split}&\tilde{\cal F}^a\equiv
(2\eta_a+\xi^a_{nn})f+(\xi^a_nx^n-\xi^a)
\frac{f'}{x}-\frac1{x^4}\left( {x^a\xi^nx^n}+
2\mu^{an}x^n+x^a\mu^{nn}\right)(x^2f''-xf')\\&
-\frac1{x^6}{x^a}\mu^{mn}x^mx^n(x^3f'''-3x^2f''+3x f')+
2\mu^{an}x^n\frac{V'}x,\end{split}\la{mmmu3}\\
x^2f\eta_{nn}+ \eta_nx^nxf'+(\xi^nx^n+\mu^{nn})x\tilde V'+
\frac1{x^2}\mu^{mn}x^mx^n(x^2V''-x\tilde V')=0. \la{mmmu4}
\end{gather}
  where $f'=\frac{\p f}{\p x}, \ \xi^a_n=\frac{\p \xi^a}{\p x_n}$, etc.

Thus to classify Hamiltonians (\ref{H}) admitting second order
integrals of motion (\ref{Q}) it is necessary to find all
inequivalent solutions of rather complicated system
(\ref{mmmu0})--(\ref{mmmu4}). Moreover, we will see that equation
(\ref{mmmu4}) can be deduced from the remaining ones. The presented
system is overdetermined and includes 19 equations for 12 unknown
functions $\mu^{ab}, \xi^a, \eta, f$ and $\tilde V$.

\section{Discussion of the determining equations}
The autonomous subsystem (\ref{mmmu0}) defines a conformal Killing tensor.
Its general solution is a linear combination of the following tensors
(see, e.g., \cite{Kil})
\begin{gather}\begin{split}&\mu^{ab}_1=\delta^{ab}\varphi_1(x)+
k (x^ax^b-\delta^{ab}x^2),\\&
\mu^{ab}_2=\lambda^a x^b+\lambda^b x^a+\delta^{ab}\lambda^c x^c\varphi_2(x),
\\&
\mu^{ab}_3=(x^a\varepsilon^{bcd}+x^b\varepsilon^{acd}) x^c\lambda^d,
\\& \mu^{ab}_4=(x^a\lambda^b+x^b\lambda^a)x^2-4x^ax^b\lambda^c x^c+
\delta^{ab}
\lambda^c x^c\varphi_3(x),
\\& \mu^{ab}_5=\lambda^{ab}+\delta^{ab}\lambda^{cd}x^cx^d\varphi_4(x),
\\&
\mu^{ab}_6=(\varepsilon^{acd}\lambda^{cb}+ \varepsilon^{bcd}
\lambda^{ca})x^d, \\&
\mu^{ab}_7=\lambda^{ab}x^2-(x^2\lambda^{bc}+x^b\lambda^{ac})x^c+
\delta^{ab}\lambda^{cd}x^cx^d\varphi_5(x),
\\&\mu^{ab}_8= 2(x^a\varepsilon^{bcd} +x^b\varepsilon^{acd})
\lambda^{dn}x^cx^n- (\varepsilon^{ack}\lambda^{bk}+
\varepsilon^{bck}\lambda^{ak})x^cx^2,\\&
\mu^{ab}_9=\lambda^{ab}x^4-2(x^a\lambda^{bc}+x^b\lambda^{ac})x^cx^2+
(4x^ax^b+k\delta^{ab}x^2)\lambda^{cd}x^cx^d+\delta^{ab}\lambda^{cd}x^cx^d
\varphi_6(x)\end{split}\la{mmu}
\end{gather}
where $\lambda^{ab}=\lambda^{ba}, \lambda^a $  are arbitrary
parameters, and $\varphi_1,...,\varphi_6$ are arbitrary functions of
$x$.

The next step is to solve the remaining equations
(\ref{mmmu2})--(\ref{mmmu4}) with $\mu^{ab}$ being linear
combinations of tensors (\ref{mmu}).
  Fortunately, this huge problem can be reduced to a series of relatively
  simple subproblems corresponding to particular linear combinations of
  these tensors.

Let us specify such linear combinations of tensors (\ref{mmu})  which should be considered separately. They should include the terms with the same transformation properties w.r.t. the rotation group. The tensors $\mu^{ab}_1, <\mu^{ab}_2,...,\mu^{ab}_4>$ and $<\mu^{ab}_5,...,\mu^{ab}_9>$ generate scalar, vector and tensor  integrals of motion correspondingly. Separating scalars, vectors and tensors with the same parities, we can specify the following non-equivalent versions of $\mu^{ab}$:
\begin{gather}\la{mue1}\mu^{ab}=\mu^{ab}_1\end{gather}
for scalar integrals of motion,
\begin{gather}\la{mue2}\mu^{ab}=\mu^{ab}_3\end{gather}
for pseudovector integrals of motion,
\begin{gather}\la{mue3}\mu^{ab}=\nu\mu^{ab}_2+\lambda\mu^{ab}_4\end{gather}
for vector integrals of motion,
\begin{gather}\la{mue4}\mu^{ab}=\nu\mu^{ab}_5+\omega\mu^{ab}_7 +\lambda\mu^{ab}_9\end{gather}
for pseudotensor integrals of motion, and
\begin{gather}\la{mue5}\mu^{ab}=\nu\mu^{ab}_6+\lambda\mu^{ab}_8 \end{gather}
for tensor integrals of motion, were $\nu, \ \lambda$ and $\omega$
are arbitrary parameters.

The linear combinations of  arbitrary functions appearing in (\ref{mue3}) and (\ref{mue4}) should be treated as new arbitrary functions.

The next subsystem of the determining equation, i.e.,  (\ref{mmmu1}), is compatible iff:
\begin{gather}\la{cc}(\mu^{nn}_a+2\mu^{na}_n)\mu^{bn}x_n=(\mu^{nn}_b+2\mu^{na}_n)\mu^{bn}x_n.\end{gather}

Then, comparing $\p_a{\cal F}^a$ with $x_ax_b{\cal F}^{ab}$ we find the following differential consequence of (\ref{mmmu1}) and (\ref{mmmu2}):
\[ 2(\xi^n_n-\mu^{kn}_{kn})f=3(\xi^nx_n-\mu^{kn}_{k}x_k)\frac{f'}x\]
which is compatible with (\ref{mmmu1}) and (\ref{mmmu2}) in two cases: either
\begin{gather}\la{cc1}\xi^n=\mu^{kn}_k\end{gather}
or the vector $\tilde \xi^n=\xi^n-\mu^{kn}_k$ satisfies the following
condition
\begin{gather*}(\tilde\xi^a_b+\tilde\xi^b_a)f=\delta_{ab}
\tilde\xi^n_n\frac{f'}x\end{gather*} which is the necessary condition for
coefficients of the first order integrals of motion \cite{NZ}.
Since such integrals of motion had been already classified in \cite{NZ},
we will set $\tilde \xi^a=0$, i.e., impose the condition (\ref{cc1}) on
coefficients $\xi^n$.

Considering  differential consequence of (\ref{mmmu2}) and
 (\ref{mmmu3}) in the forms $\p_a\p_b{\cal F}^{ab}=0$ and
 $\p_a\tilde{\cal F}^a=0$ we obtain equation (\ref{mmmu4}). So the
 latter equation is a consequence of (\ref{mmmu2}) and
 (\ref{mmmu3}) and can be omitted.

Thus the problem of classification of rotationally invariant PDM
systems admitting second order integrals of motion is reduced to
search for inequivalent solutions of equations (\ref{mmmu1}),
(\ref{mmmu2}) and (\ref{mmmu3}) for unknowns $f$, $\tilde V$,
$\xi^a$ and $\eta$ for all versions of functions $\mu^{ab}$
enumerated in (\ref{mue1})--(\ref{mue5}). The corresponding
calculations are outlined in Appendix, while the classification
results are presented in the following section.

Let us note that whenever condition (\ref{cc1}) is satisfied and
functions $f, \tilde V$ and $\varphi_1, \varphi_2,...,\varphi_6,
\eta$ in (\ref{H}) and (\ref{Q}), (\ref{mmu}) are real, both
Hamiltonians $\hat H$ and second order integrals of motion $Q$ are
formally self-adjoint on the standard $L_2$ space with scalar
product \beg\la{sp}<\psi_1|\psi_2>=\int_{{\cal M}}\bar
\psi_1\psi_2d^3x.\end{gather}  Just using the standard scalar
product (\ref{sp}) is one of the  main points of our approach.

\section{Classification results}
\subsection{Equivalence transformations}

It was indicated in \cite{NZ} that the equivalence group of equation
(\ref{se}) is nothing but C(3), i.e., the conformal group in $3d$
Euclidean space. It means that acting on dependent and independent
by transformations belonging to C(3), we do not change the generic
form of Hamiltonian (\ref{H}) although functions $f$ and $\tilde V$
can be changed.

However, since we suppose that equation (\ref{se}) is rotationally
invariant, we should restrict ourselves to such equivalence
transformations which keep this invariance, i.e., are either pure
rotations or transformations commuting with $J_a$ (\ref{J}). In
other words, the equivalence transformations for the considered
class of equations are reduced to products of rotations and scalings
of independent variables
\begin{gather}\la{et}x_a\to R_{ab}x_b, \quad x_a\to \omega
x_a,\end{gather} and the  inverse transformation \beg\la{IT} x_a\to
\tilde x_a=\frac{x_a}{x^2},\quad \psi({\bf x})\to \tilde x^3\psi(\tilde{\bf x})\end{gather} where $R_{ab}$ is a rotation matrix and
$\omega$ is an arbitrary real parameter. In addition, we will define
Hamiltonians $H$ up to multiplication by a real parameter $\omega$
and up to a constant  shift of potential $\tilde V$. Thus the
equivalence transformations (\ref{et}) and (\ref{IT}) will be
extended by the following changes \beg\la{et2}H\to \omega H,\quad
\tilde V\to \tilde V+C\end{gather} where $\omega$ and $C$ are real
constants.

In the following we present the Hamiltonians and the corresponding
second order integrals of motion obtained by solving the system
(\ref{mmmu0})--(\ref{mmmu4}), see Appendix for calculation details.
These solutions are defined up to equivalence transformations
(\ref{et}), (\ref{IT}) and (\ref{et2}).

The presentation (\ref{H}) for Hamiltonians is compact and convenient for our classification procedure. However there exist another and physically motivated representation \cite{Roz}, \cite{mora}, which is equivalent to (\ref{H}):
   \begin{gather}\la{last}  H=m^r p_a m^{-1-2r}p_am^r+ V\end{gather}
   where $m=\frac1f$ and $r$  is the ambiguity parameter of the kinetic energy term \cite{Roz}. We will present the found Hamiltonians in the form (\ref{last}) with $r=-1/2$, i.e.,
   \begin{gather}\la{last2}  H=f^\frac12 p^2 f^\frac12 + V.\end{gather}

   It happens that just representation (\ref{last2}) corresponds to the
   most compact forms of the mass and potential terms. The related equation
   (\ref{se}) should be rewritten without hats and tildes:
\begin{gather}\la{se3}
    H \psi=E \psi.
\end{gather}
   Notice that potentials $V$ and $\tilde V$ are connected by the following relation:
   \begin{gather}V=\tilde V+\frac{f'}x+\frac{f''}2-\frac{f'^2}{4f}.
   \la{last3}\end{gather}

  Nonequivalent Hamiltonians (\ref{last2}) admitting second order integrals
  of motion are presented in the following subsections.
  \subsection{Vector integrals of motion}
 In Section 3 three classes of second order integrals of motion had been
 indicated,
 i.e., scalar, vector and tensor ones. It is shown in Appendix  that the scalar integrals of motion are linear combinations of Hamiltonian and squared orbital momentum and so can be treated as trivial.  Thus we start with vector symmetries.

   In accordance with the
analysis presented in section 3 it is possible to specify two
tensors $\mu^{ab}$ which can generate vector integrals of motion,
i.e.,
 \begin{gather}\la{V1}\mu^{ab}=\mu^{ab}_3=(x^a\varepsilon^{bcd}+
 x^b\varepsilon^{acd}) x^c\lambda^d_3 \end{gather}
and a linear combination of tensors $\mu^{ab}_2$ and $\mu^{ab}_4$:
\begin{gather}\la{V2}\begin{split}&
\mu^{ab}=\nu\mu^{ab}_2+\mu\mu^{ab}_4=\nu(\lambda^a x^b+\lambda^b x^a-2\delta^{ab}\lambda^c x^c)\\&+\mu((x^a\lambda^b+x^b\lambda^a)x^2-4x^ax^b\lambda^c x^c+2\delta^{ab}
x^2\lambda^c x^c)+\delta^{ab}
\lambda^c x^c\varphi(x)\end{split}\end{gather}
where we use the arbitrariness of $\varphi$ to obtain a convenient realization for $\mu^{ab}$.

Versions (\ref{V1}) and (\ref{V2}) should be considered separately, since the related  integrals of motion have different parities w.r.t. the space inversion.

It is shown in Appendix that integrals of motion corresponding to
(\ref{V1}) are admitted only by the systems specified in
(\ref{fV1}). Moreover, these integrals of motion are nothing but
polynomials in the first order symmetries (\ref{J}) and (\ref{I1}).

Considering integrals of motion corresponding to (\ref{V2}) we can a priory restrict ourselves to the following values of parameters $\alpha$ and $\mu$:
\begin{gather}\la{par1}\nu=1,\quad \mu=0;\\\la{par2}\nu=\mu=1;\\\la{par3}\nu=-\mu=1.\end{gather}
Then integrals of motion corresponding to arbitrary $\alpha$ and
$\mu$ can be obtained by scaling and inversions of independent
variables $x_a$, i.e, by products of  transformations (\ref{et}) and
(\ref{IT}).

Substituting (\ref{V2}) into (\ref{cc}) we obtain the following equation for function $\varphi$:
\[x(\nu+\lambda x^2)\varphi'=\varphi^2+(3\lambda x^2-\nu)\varphi,\]
whose solutions are:
\begin{gather}\la{phi}\varphi=0, \quad \text {and} \quad
\varphi=\frac{(\nu+\mu x^2)^2}{\nu-2\kappa x-\mu x^2}\end{gather}
where $\kappa$ is an integration constant. Then, substituting
(\ref{V2}) and (\ref{phi}) into equations (\ref{mmmu1}),
(\ref{mmmu2}) and going over values of parameters  $\nu$ and $\mu$
specified in (\ref{par1})--(\ref{par3}), we find admissible
functions $f$. The corresponding potentials can be found solving the
remaining determining equation, i.e., (\ref{mmmu3}). The results of
these calculations (whose details can be found in Appendix) are
presented in Table 1.

In the table all non-equivalent mass and potentials are presented which give rise to superintegrable systems admitting vector integrals of motion. However, this list can be added by the systems whose masses and potentials are specified in (\ref{fV1})--(\ref{fV4}). The latter systems also admit second order integrals of motion, which are products of their first order symmetries. Such symmetries are apparent and will not be discussed here.

 {\small \begin{center}Table 1. Functions $f$ and $V$ specifying non-equivalent Hamiltonians (\ref{last2}) and the corresponding  vector integrals of motion.\end{center}
\begin{tabular}{|c|c|c|c|c|c|}
\hline
No&$f$&$V$&\text{Integrals of motion}&$\begin{array}{c}\text{Solution}\\\text{approach}\end{array}$&$\begin{array}{c}\text{Effective}\\\text{potentials}\end{array}$\\
\hline
&&&&&\\
1.&$x$&$\alpha x$&$Q_a=\{p_b,J_{ab}\}+\frac12\{H,\frac{x_a}x\}$&$\begin{array}{c}
\text{direct}\\\text{or two-step}\end{array}$&$\bea{c}\text{3d oscillator} \\\text{or Coulomb}\eea$\\
2.&$x^4$&$\alpha x$&$Q_a=\{K_b,J_{ab}\}- \alpha x^a$&$\begin{array}{c}
\text{direct}\\\text{or two-step}\end{array}$&$\bea{c}\text{Coulomb}\\\text{or 3d oscillator}\eea$\\
3.&$x(x-1)^2$&$\frac{\alpha x}{(x+1)^2}$&$Q_a=
\{J_{ab},N^+_b\}+\frac12\{H,\frac{x_a}x\}
$&$\begin{array}{c}
\text{direct}\\\text{or two-step}\end{array}$&$\bea{c} \text{Eckart}\\\bea{c}\text{or hyperbolic}\\\text{P\"oschl-Teller}\eea
\eea$\\
4.&$x(x+1)^2$&$\frac{\alpha x}{(x-1)^2}$&$\begin{array}{l}Q_a=
\{J_{ab},N^+_b\}+\frac12\{H,\frac{x_a}x\}\end{array}$
&$\begin{array}{c}
\text{direct}\\\text{or two-step}\end{array}$&$\bea{c}\text{ Eckart}\\\text{or trigonometric}\\
\text{P\"oschl-Teller}\eea$\\
5.&$(1+x^2)^2$&$\frac{\alpha (1-x^2)}{x}$&$Q_a=
\{J_{ab},N^-_b\} -\alpha\frac{x^a}x$&direct&$\bea{l}\text{trigonometric}\\
\text{Rosen-Morse}\eea$
\\
6.&$(1-x^2)^2$&$\frac{\alpha (1+x^2)}{x}$&$Q_a=
\{J_{ab}N^+_b\}- \alpha\frac{x^a}x$&direct&$\bea{c}\text{ Eckart}
\eea$\\&&&&&\\
7.&$\frac{x}{ x+1}$&$\frac{\alpha x}{
x+1}$&$Q_a=\{J_{ab},p_b\}+\frac12\{H,\frac{x_a}x\}$ &two-step&$\bea{c}
\text{Coulomb }\eea$\\&&&&&\\
8. &$\frac{x}{x-1}$&$\frac{\alpha x}{ x-1}$&$\bea{l}Q_a=\{J_{ab},p_b\}+\frac12\{H,\frac{x_a}x\}\eea$&two-step&Coulomb \\
9.&$\bea{l}\frac{(x^2-1)^2x}{x^2-2\kappa x+1}\eea$&$\frac{\alpha x}{x^2-2\kappa x+1}$&$
\begin{array}{l}Q_a=
\{J_{ab},N^+_b\}+\frac12\{H,\frac{x_a}x\}\end{array}$&two-step&
$\bea{l}\text { Eccart}
\eea$\\
10.&$\frac{(x^2+1)^2x}{x^2-2\kappa x-1}$& $\frac{\alpha
x}{x^2-2\kappa x-1}$&$\begin{array}{l}Q_a=
\{J_{ab},N^-_b\}+\frac12\{H,\frac{x_a}x\}
\end{array}$&two-step&$\bea{l}\text{trigonometric}\\
\text{Rosen-Morse}\eea$
\\
\hline
\end{tabular}}

\vspace{2mm}

Here $\alpha$ and $
\kappa\neq \pm1$ are arbitrary constants, the symbol $\{.,.\}$ denotes anticommutator,\\
$J_{ab}=\varepsilon_{abc}J_c$, while operators $K_b, N^\pm_b$ and  $ J_a$
are defined in equations (\ref{I2})--(\ref{I4}) and (\ref{J}).

All presented   systems are shape invariant. More exactly,
 this property is possessed by the corresponding radial equation. The way to obtain the radial equations with the indicated shape invariant potentials is described algorithmically in section 6.1. For the cases enumerated in Items 1-4 we have two-fold shape invariance when two different
 superpotentials can be used to factorize the radial equation.
 The types of  radial potentials are indicated in the last column.

The list of potentials and mass term presented in Table 1
is
completed up to equivalence transformations (\ref{et}), (\ref{IT})
and (\ref{et2}). Using these transformations, it is possible to
propagate the obtained systems to families of equivalent ones. For
example, starting with the system specified in Item 1 and making the
inverse transformation (\ref{IT}), we can construct the following
Hamiltonian and the related integrals of motion:
\begin{gather}\begin{split}
&H=x^\frac32p^2x^\frac32+\frac{\alpha}{ x},\\&
Q_a=J_{ab}K_b+K_bJ_{ab}+\frac12\left\lbrace H,\frac{x_a}x\right\rbrace.
\end{split}\la{sys2}\end{gather}

The same trick can be made with the systems specified in Items 2, 7 and 8 while the remaining systems are invariant w.r.t. the inversion transformation up to signs of Hamiltonian or parameter $\alpha$.
\subsection{Tensor integrals of motion}
 Consider now determining equations (\ref{mmmu1})--(\ref{mmmu4}) with
 tensor $\mu^{ab}$ given by formula (\ref{mue4}), i.e.,
 \begin{gather}\la{a41}\begin{split}&\mu^{ab}=\nu\lambda^{ab}+
 \omega\left(\lambda^{ab}x^2-(x^2\lambda^{bc}+x^b\lambda^{ac})x^c
 -2\delta^{ab}\lambda^{cd}x^cx^d\right)\\& +
 \mu\left(\lambda^{ab}x^4-2(x^a\lambda^{bc}+x^b\lambda^{ac})x^cx^2+
 4x^ax^b\lambda^{cd}x^cx^d\right)+\delta^{ab}\lambda^{cd}x^cx^d\varphi(x).
 \end{split}\end{gather}

 The term multiplied by  $\omega$ is not essential since
 it corresponds to integrals of motion proportional to $J^aJ^b$,
 which are accepted by any Hamiltonian (\ref{last2}) thanks to its
 rotational invariance. However, the presence of this term helps to write
 some of integrals of motion in a more compact form.

 Like in previous section it is sufficient to restrict ourselves to the
 values of parameters $\nu$ and $\mu$ fixed in equations (\ref{par1}),
 (\ref{par2}) and (\ref{par3}). Solving the corresponding equations
 (\ref{mmmu1})-(\ref{mmmu3}) we find all systems admitting pseudotensor
 integrals of motion, see Appendix for calculation details. The classification results are
 presented in Table 2.
{\small
\begin{center}Table 2. Functions $f$ and $V$ specifying non-equivalent
Hamiltonians (\ref{last2}) which admit second order pseudotensor
integrals of motion.\end{center}

\begin{tabular}{|c|c|c|c|c|c|}
\hline
No&$f$&$V$&\text{Integrals of motion}&$\begin{array}{c}\text{Solution}\\\text{approach}\end{array}$&$\bea{c}\text{Effective}\\\text{radial}\\\text{potential}\eea$\\
\hline
&&&&&\\
1.&$\frac{1}{x^2}$&$\frac\alpha{x^2}$&$\bea{c}Q_{ab}=p_ap_b-\\-
\frac12\left\{x_ax_b,H+\frac1{x^4}\right\}\eea
$&$\begin{array}{c}
\text{direct}\\\text{or two-step}\end{array}$&$\bea{c}\text{Coulomb}\\\text{or 3d oscillator}\eea$\\
2.&$x^4$&$-\frac\alpha{x^2}$&$Q_{ab}=K_aK_b- \alpha\frac{x^ax^b}{x^4}$&$\begin{array}{c}
\text{direct}\\\text{or two-step}\end{array}$&$\bea{c}\text{3d oscillator}\\\text{or Coulomb}\eea$\\
3.&$(x^2-1)^2$&$\frac{\alpha x^2}{(x^2+1)^2}$&$Q_{ab}=\{N^+_a,N^+_b\}+\frac{2\alpha x^ax^b}{(x^2+1)^2}$&$\begin{array}{c}
\text{direct}\\\text{or two-step}\end{array}$&$\bea{c}\text{Eckart}\\\text{or hyperbolic}\\
\text{P\"oschl-Teller}\eea$\\
4.&$(x^2+1)^2$&$\frac{\alpha x^2}{(x^2-1)^2}$&$Q_{ab}=\{N^-_a,N^-_b\}+\frac{2\alpha x^ax^b}{(x^2-1)^2}$&$\begin{array}{c}
\text{direct}\\\text{or two-step}\end{array}$&$\bea{c}\text{Eckart}\\\text{or trigonometric}\\
\text{P\"oschl-Teller}\eea$\\
5.&$\frac{(x^4-1)^2}{x^2}$&$\frac{\alpha(x^4+1)}{x^2}$&
$\bea{c}Q_{ab}=K_aK_b+p_ap_b-\\-\frac12\left\lbrace p_c(1+x^4)p_c+\alpha,\frac{x_ax_b}{x^2}\right\rbrace\eea$&direct&$\bea{c}\text{ Eckart}\\\eea$\\
6.&$\frac{(x^4+1)^2}{x^2}$&$\frac{\alpha(x^4-1)}{x^2}$&
$\bea{c}Q_{ab}=K_aK_b-p_ap_b-\\-\frac12\left\lbrace p_c(x^4- 1)p_c+\alpha,\frac{x_ax_b}{x^2}\right\rbrace\eea$&direct&$\bea{c}
\text{trigonometric}\\\text{Rosen-Morse}\eea$\\
7.&$\frac1{x^2+1}$&$\frac{\alpha }{x^2+1}$&$\bea{c}Q_{ab}=p_ap_b-\\-
\frac12\left\{x_ax_b,H+\frac1{(x^2+1)^2}\right\}\eea$&two-step&$\bea{c}\text{3d oscillator}\eea$\\
8.&$\frac1{x^2-1}$&$\frac{\alpha }{x^2-1}$&$\bea{c}Q_{ab}=p_ap_b-\\-
\frac12\left\{x_ax_b,H+\frac1{(x^2-1)^2}\right\}\eea$&two-step&$\bea{c}\text{3d oscillator}\eea$\\
9.&$\frac{(x^4-1)^2}{x^4-2\kappa x^2+1}$&$\frac{\alpha x^2}{x^4-2\kappa x^2+1}$&$\begin{array}{c}
Q_{ab}=K_aK_b +p_ap_b\\-\frac12\left\lbrace H+6\kappa+\right.\\\left.+p_c(x^4+1)p_c,\frac{x_ax_b}{x^2}\right\rbrace\end{array}
$&two-step&$\bea{c}\text{ Eckart}\eea$\\
10.&$\frac{(x^4+1)^2}{x^4-2\kappa x^2-1}$&$\frac{\alpha x^2}{x^4-2\kappa x^2-1}$&$
\begin{array}{c}
Q_{ab}=K_aK_b-p_ap_b-\\-\frac12\left\lbrace H+6\kappa+\right.\\ \left.+p_c(x^4-1)p_c,\frac{x_ax_b}{x^2}\right\rbrace\end{array}$&two-step&$\bea{c}
\text{trigonometric}\\\text{Rosen-Morse}\eea$\\&&&&&
\\
\hline \end{tabular}}

\section{Shape invariance and exact solutions}
All systems presented in Tables 1 and 2 are maximally superintegrable and can be solved exactly. In addition, all of them appear to be shape invariant and so can be solved using tools of SUSY quantum mechanics. Moreover, some of the presented systems are characterized by the multiple shape invariance, i.e., they can be solved using more then one superpotential.
\subsection{Two strategies in construction of exact solutions}

Let us consider  equations  (\ref{se3}) where $H$ are hamiltonians (\ref{last2}) whose mass and potential terms are specified in the presented tables. We will search for square integrable solutions of these systems vanishing at $x=0$.

 First let us transform (\ref{se3}) to the following equivalent form
 \begin{equation}\la{se1}
   \tilde H \Psi=E \Psi,
\end{equation}
where
  \begin{gather}\la{Hef} \tilde H=\sqrt{f}H \frac{1}{\sqrt{f}}=fp^2+V,
  \quad \Psi=\sqrt{f}\psi. \end{gather}

  Then, introducing spherical variables and expanding solutions via spherical functions $Y^l_m$
 \begin{gather}\la{rv}\Psi=\frac1x\sum_{l,m}\phi_{lm}(x)Y^l_m\end{gather}
we obtain the following equation for radial functions:
\begin{gather}\la{re}-f\frac{\p^2\phi_{lm}}{\p x^2}+\left(\frac{fl(l+1)}{x^2}+V \right)\phi_{lm}=E\phi_{lm}.\end{gather}

We will search for normalizable solutions of equations (\ref{se1})
and (\ref{re2}). In accordance with (\ref{sp}) and (\ref{Hef}) the
corresponding scalar products look as follows:
\beg\la{sp2}<\Psi_1|\Psi_2>=\int_{{\cal M}}\bar
\Psi_1\Psi_2f^{-1}d^3x.\end{gather} and
\beg\la{sp3}<\phi_{lm}^{(1)}|\phi_{lm}^{(2)}>=\int_{0}^R\bar
\phi_{lm}^{(1)}\phi_{lm}^{(2)}f^{-1}dx\end{gather} respectively,
where
 the integration limit  $R$ is equal to 1 or $R\to\infty$ depending on a
 concrete problem. All solutions presented in the following text are
 normalizable with respect to the scalar product (\ref{sp3}) with
 $R=\to\infty$.

 Let us present two
possible ways to solve equation (\ref{re}). They can be treated as particular cases of Liouville transformation (refer to \cite{olver} for definitions) and include commonly known steps. But it is necessary to fix them as concrete algorithms to obtain shape invariant potentials presented in the tables.

The first way (which we
call direct) includes consequent changes of independent and
dependent variables: \beg \phi_{lm}\to \Phi_{lm}=
f^{\frac14}\phi_{lm}, \ \frac{\p}{\p x}\to f^{\frac14} \frac{\p}{\p
x} f^{-\frac14}= \frac{\p}{\p
x}+\frac{f'}{4f}\la{change}\end{gather} and then \beg \la{change3}
x\to y(x), \end{gather} where $y$ solves the equation $\frac{\p
y}{\p x}=\frac1{\sqrt{f}}$.  As a result equation (\ref{rv}) will be
reduced to a more customary form
\begin{gather}\la{re2}-\frac{\p^2\Phi_{lm}}{\p y^2}+\tilde V  \Phi_{lm}=
E\Phi_{lm}\end{gather} where $\tilde V$ is an effective potential
\beg\la{po}\tilde V=V+f\left({\frac{l(l+1)}{x^2}} -\left(\frac{
f'}{4 f}\right)^2-\left(\frac{ f'}{4 f}\right)'\right),\quad
x=x(y).\end{gather}

  Equations (\ref{se1}), (\ref{Hef})
with functions $f$ and $V$ specified in Items 1--6 of both Tables 1
and 2 can be effective solved using the presented reduction to
radial equation (\ref{re2}). All the corresponding potentials (\ref{po}) appears to be shape invariant, and just these potentials are indicated in the fifth columns of the tables. The related  equations
(\ref{re2}) are shape invariant too and can be solved using
the SUSY routine.

However, if we apply the direct approach to  the remaining systems (indicated in Items 7 -- 10 of both  tables), we come to equations  (\ref{re2}) which are not shape invariant and are hardly solvable, if at all.  To solve these systems we need a more sophisticated  procedure which we call two-step approach. To apply it we multiply (\ref{re}) by $\alpha V^{-1}$
and obtain the following equation:
\begin{gather}\la{re3}-\tilde f\frac{\p^2\phi_{lm}}{\p x^2}+\left(\frac{\tilde f l(l+1)}{x^2}+\tilde V \right)\phi_{lm}={\cal E}\phi_{lm}\end{gather}
where $\tilde f=\frac{\alpha f}{ V},\ \tilde V =-\frac{\alpha E}V$ and $ {\cal E}=-\alpha$ .
Then treating $\cal E$ as an  eigenvalue and solving equation (\ref{re3}) we can find $\alpha$ as a function of $E$, which defines admissible energy values at least implicitly. To do it it is convenient to make changes (\ref{change}) and (\ref{change3}) where $f\to \tilde f$.

The presented trick with a formal changing the roles of constants $\alpha$ and $E$ is well known.  Our
point is that {\it any of the presented superintegrable systems can
be effective solved using either the direct approach presented in
equations (\ref{Hef})--(\ref{po}), or the two-step approach}. Moreover, some of the presented systems can be solved using both the direct and two-step approaches, as indicated in the fourth columns of Table 1 and 2. In all cases we obtain shape
invariant effective potentials and can use tools of SUSY quantum
mechanics.

\subsection{A system with two-fold shape invariance}
Let us apply the presented algorithms to selected superintegrable systems. We start with the following hamiltonian
\beg H=\frac1xp^2\frac1x+\frac{\alpha}{x^2}\la{h11}\end{gather}
which corresponds to functions $f$ and $V$ specified in the first item of Table 2. The corresponding radial equation (\ref{re}) takes the following form
\begin{gather}\la{re4}-\frac{1}{x^2}\frac{\p^2\phi_{lm}}{\p x^2}+\left(\frac{l(l+1)}{x^4}+\frac{\alpha}{x^2} \right)\phi_{lm}=E\phi_{lm}.\end{gather}

Equation (\ref{re4}) can be effectively solved using the direct method presented in the previous section. Making changes (\ref{change}), (\ref{change3}) with  $y=\frac{x^2}2$  we obtain the following version of equation (\ref{re2}):
\begin{gather}\la{rey}{\cal H}_\mu\Phi_{\mu m}\equiv\left(-\frac{\p^2}{\p y^2}+\frac{\mu(\mu
+1)}{y^2}+\frac{\alpha}{2y} \right)\Phi_{\mu m}=E\Phi_{\mu m}\end{gather}
where
\beg\mu=\frac{l}2-\frac14\la{mu}.\end{gather}

Up to the meaning of parameter $\mu$  equation (\ref{rey}) formally coincides with the radial equation for Hydrogen atom, provided $\alpha<0$. Hamiltonian ${\cal H}_\mu$ is shape invariant, so we can construct exact solutions using the following procedure.

Let us set
$\alpha=-2\nu^2$ and construct the corresponding solutions.  To do that it is possible to use the nice symmetry of the Hydrogen atom called shape invariance. We will present this routine procedure since it can be applied for all systems specified in Tables 1,2. In addition, there is some fine points connected with the non-standard definition of parameter $\mu$.

Like in the case of standard Hydrogen atom, Hamiltonian ${\cal H}_\mu$ can be factorized:
\begin{gather}\la{fact}{\cal H}_\mu=a_\mu^+ a_\mu^- +c_\mu\end{gather}
where
\begin{gather}\la{a}a^-_\mu=\frac{\p}{\p y}+ \quad a^+_\mu=-\frac{\p}{\p y}+W_\mu, \quad c_\mu=\frac{\nu^4}{4(\mu+1)^2},\end{gather}
and $W_\mu =\frac{\nu^2}{2(\mu+1)}- \frac{\mu+1}y$ is a superpotential.

Hamiltonian ${\cal H}_\mu$ is shape invariant since ${\cal H}^+_\mu=a^-_\mu a^+_\mu={\cal H}_{\mu+1}+c_\mu-c_{\mu+1}$. Thus equation (\ref{rey}) can be integrated using the standard tools of supersymmetric quantum mechanics, see, e.g., \cite{khare}.

The ground state $\Phi^0_{lm}$ solves the first order equation $a^-_\mu \Phi^0_{lm}=0$ and is given by the following formula:
\begin{gather*}\Phi^0_{\mu}=C_0y^{\mu+1}e^{-\frac{\nu^2 y}{(\mu+1)}} \end{gather*}
where subindices $l,m$ are omitted. The corresponding value of $E$
in (\ref{rey}) is equal to $-c_\mu$.

The standard expressions for the first, second and $n^{\text th}$ exited state and the corresponding energies are:
\begin{gather}\la{1}\Phi^1_{\mu}=a^+_\mu\Phi^1_{\mu+1},\quad E_1=\frac{\nu^4}{4(\mu+2)^2},\\
\la{2}\Phi^2_{\mu}=a^+_\mu a^+_{\mu+1}\Phi^0_{\mu+2},\quad E_2=\frac{\nu^4}{4(\mu+3)^2}
\end{gather}
and
\begin{gather}\la{n}\Phi^n_{\mu}=a^+_\mu a^+_{\mu+1}...a^+_{\mu+n-1}\Phi^{0}_{\mu+n},\end{gather}\begin{gather}
E_n=-\frac{\nu^4}{4(\mu+n+1)^2}=-\frac{\alpha^2}{(4n+2l+3)^2}\la{En}\end{gather}
correspondingly. The explicit expression for the related radial function $\phi^n_{lm}$  can be found by substituting (\ref{n}) into (\ref{change}):
\begin{gather}\la{phi1}\phi_{lm}^n=C_{lm}^{n} z^\frac{2l+{3}}4\exp\left(-\frac{z}2\right){\cal F}\left(-n,l+\frac32,z\right)\end{gather}
where $\cal F$ is the confluent hypergeometric function and $z=\frac{-\alpha x^2}{(4n+2l+3)}$.

 Thus, exploiting the shape invariance of equation  we find the admissible
 eigenvalues and the corresponding eigenvectors. It was done in analogy
 with the standard Coulomb problem. However, in fact we were supposed to
 generalize the standard approach in the following point. Usually, to find
 the $n^{\text th}$ exited state we act by creation operators on the
 ground state, which is an eigenvector of the squared orbital momentum.
 In our case already the first exited state (\ref{1}) is expressed via
 $\Phi^0_{\mu+1}$ {\it which does not solve equation (\ref{rey}) with any
 admissible value of $\mu$ presented in (\ref{mu}).} More exactly,
 let $\mu$ solves equation (\ref{mu}) with some $l$, and $\mu'=\mu+1$
 solves (\ref{mu}) with   $l'$. Then $l$ and $l'$ cannot be integers
 simultaneously.

 In other words, to construct the first exited state in fact we use a
 virtual ground state. The same is true for all odd states. However,
 they solve the auxiliary equation of type (\ref{rey}) with $\mu\to \mu-1$,
 which is a superpartner of (\ref{rey}).

Thus the odd and even states are logically separated.  It is
possible to separate them also formally using the approach discussed
in \cite{BDN}

Let us show that equation (\ref{re4}) can be solved also using the two-step approach discussed in section 6.1. Indeed, multiplying it by $x^2$ we immediately come to the following equation:
\begin{gather}\la{re5}{\cal H}_l\phi_{lm}\equiv \left(-\frac{\p^2}{\p x^2}+\frac{l(l+1)}{x^2}+\frac{\omega^2}4{x^2} \right)\phi_{lm}={\cal E}\phi_{lm}\end{gather}
where we denote $-E=\frac{\omega^2}4$ and $-\alpha={\cal E}$.

In other words, we come to another shape invariant system which is nothing
but 3$d$ isotropic harmonic oscillator. Using again the standard technics of SUSY quantum mechanics \cite{khare}   we can find admissible eigenvalues $\cal E$ and the corresponding state vectors in the following form:
\begin{gather}
{\cal E}=\omega(2n+l+3/2)\la{Enn}\end{gather}
and
\begin{gather}\la{philm} \phi^n_{lm}=C^n_{lm}x^{l+1}e^{\frac{-\omega x^2} 4}L_n^{l+\frac12}\left(\frac{\omega x^2}{2}\right)\end{gather}
where
$L_n^{l+\frac12}\left(\frac{\omega x^2}{2}\right)$ are Laguerre polynomials.

As expected, formulae (\ref{Enn}) and (\ref{n}) are in perfect agreement: solving (\ref{Enn}) for $\omega$ we find that $\frac{\omega^2}{4}=E_n$.

We see that hamiltonian (\ref{h11}) has rather specific properties.
Namely, it gives rise to two different shape invariant radial
equations, which formally coincide with equations for the Coulomb
and 3d oscillator problems.  Let us note that the same property,
i.e., the existence of more than one shape invariant effective
potentials, is possessed  by all systems specified in Items 1 -- 4
of both Table 1 and 2. Moreover, the system presented in Item 1 of Table 1 also presents a bridge between oscillator and Coulomb systems.
\subsection{Systems with two arbitrary parameters}
Among the Hamiltonians specified in Tables 1 and 2 there are four operators
including pairs of arbitrary parameters, namely,  $\alpha$ and $\kappa$,
see Items 9 and 10 of both tables.

Here   just these systems including pairs of arbitrary parameters are discussed. All of them are exactly solvable. Moreover, to find their solutions it is reasonable to use the two-step approach outlined in Section 6.1.

Let us start with the systems specified in Item 10 of Table 2. The corresponding Hamiltonian (\ref{Hef}) and radial equation (\ref{re}) have the following form:
\begin{gather*} H=\frac{(x^4+1)^2}{x^4-2\kappa x^2-1}p^2 +\frac{\alpha x^2}{x^4-2\kappa x^2-1}\end{gather*}
and
\begin{gather}\la{re8}\begin{split}&\left(-
\frac{(x^4+1)^2}{x^4-2\kappa x^2-1}\left(\frac{\p^2}{\p x^2}-\frac{l(l+1)}{x^2}\right)+
\frac{\alpha x^2}{x^4-2\kappa x^2-1}\right)\phi_{lm}= E\phi_{lm}\end{split}.\end{gather}

Multiplying (\ref{re8}) from the left by $\frac{x^4-2\kappa x^2-1}{x^2}$ we come to the following equation:
\begin{gather}\la{re81}\begin{split}&\left(-
\frac{(x^4+1)^2}{x^2}\left(\frac{\p^2}{\p
x^2}-\frac{l(l+1)}{x^2}\right)+\frac{\tilde\alpha(x^4-1)}{x^2}\right)
\phi_{lm}= {\cal E}\phi_{lm}\end{split}\end{gather} where
\beg\la{ev5}\tilde\alpha=-E \quad \text{and} \quad {\cal
E}=-\alpha-2\kappa E.\end{gather}

Notice that equation (\ref{re81}) with $\tilde \alpha\to \alpha$ and
${\cal E}\to E$ is needed also to find eigenvectors of the
Hamiltonian whose mass and potential terms are specified in Item 6
of Table 2.

Making transformations (\ref{change}) and (\ref{change3}) with
$f=\frac{(x^4+1)^2}{x^2}$ and $y=\frac12\arctan(x^2)$ we reduce
equation (\ref{re81}) to the following form:
\begin{gather}-\frac{\p^2\Phi_{lm}}{\p y^2} +
\left(\mu(\mu-4)\csc^2(4y)+2\tilde\alpha \cot(4y)\right)\Phi_{lm}=
\tilde{\cal E}\Phi_{lm}\la{re9}\end{gather}
where
\begin{gather}\la{EEE}\tilde {\cal E}={\cal E}+4,\quad \mu=2l+3.\end{gather}

Thus we again have equation with a shape invariant (Rosen-Morse I) potential. It is consistent provided parameters $\tilde \alpha$ and $\mu$ are positive. The corresponding eigenvalues and eigenvectors are  (see, e.g., \cite{khare})
\beg\la{ev9}\tilde{\cal E}=(\mu+4n)^2-\frac{\tilde\alpha^2}{(\mu+4n)^2},\end{gather}
and
\beg\la{psin9}\Phi_{lm}=(z^2-1)^{-\frac18(\mu+4n)}\exp(\lambda y)P_n^{\left(\frac14(\ri\lambda-\mu-4n),-\frac14(\ri\lambda+\mu+4n)
\right)}(z)
\end{gather}
where $\lambda=\frac{\tilde \alpha}{\mu+4n}$ and $z=\ri\cot(4y)$.

The initial wave functions $\phi^n_{lm}$ written in terms of the initial variable $x$ can be presented by the following expressions:
\begin{gather}\la{solo1}\phi^n_{lm}=\sqrt{\frac{x^2}{{1+x^4}}}\Phi^n_{lm}(z),\quad z=\ri\frac{1+x^4}{2x^2}.\end{gather}

Thus we find eigenvalues ${\cal E}={\cal E}_n$ and    the
corresponding state vectors $\phi^n_{lm}$ for  radial equation
(\ref{re8}). One more   effort is needed to find the explicit
expression for eigenvalues $E=E_n$, which can be found as solutions
of the system of algebraic equations (\ref{ev5}), (\ref{EEE}) and
(\ref{ev9}):
\beg\la{ev10}E_n=(2l+3+4n)^2\left(\kappa-\sqrt{\kappa^2+1
+\frac{\alpha-4}{(2l+3+4n)^2}}\ \ \right).\end{gather}

In order equation (\ref{re9}) to be consistent both parameters
$\tilde \alpha$ and $\mu$ should be positive \cite{khare}. It means
that $E_n$ should be negative, which is guaranteed if parameters
$\alpha$ and $\kappa$ satisfy the following condition:
\begin{gather*}\alpha>-5-\frac92(k-|k|)k.\end{gather*}

 In complete analogy with the above it is possible to solve
equations (\ref{se1}) for all cases specified in Items 9,10 of Table
1 and Item 9 of Table 2. To save a room we restrict ourselves to
presentation of final results.

 The energy spectrum of Hamiltonian (\ref{Hef}) with mass and potential terms fixed in Item 9 of Table 2 is given by the following formula:
 \beg\la{ev11}E_n=(2l+3+4n)^2\left(\kappa+\sqrt{\kappa^2-1
+\frac{\alpha+4}{(2l+3+4n)^2}}\ \right)\end{gather} while the
related radial state vectors are:
\beg\la{es11}\phi_{lm}^n=x\left(\frac{x^4-1}{x^2}\right)^{1-\frac{N}4}
\left(\frac{x^2+1}{x^2-1}\right)^{\frac{E_n}{4N}}
P_n^{\left(\frac{-E_n}{4N}-\frac{N}4,\frac{E_n}{4N}-\frac{N}4\right)}(z)\end{gather}
where $P_n^{(.,.)}(z)$ are Jacobi polynomials,
$z=\frac{1+x^4}{2x^2}$ and $N=2l+3+4n$. Moreover, parameters
$\alpha$ and $\kappa$ are restricted by the the following
conditions:
\begin{gather*}\alpha>5-\frac92(\kappa+|\kappa|)\kappa, \quad |\kappa|\geq1.\end{gather*}

 The
eigenvalues and radial state vectors of Hamiltonian (\ref{Hef})
specified in Item 10 of Table 1 are:
\beg\la{ev12}E_n=4(l+1+n)^2\left(\kappa-\sqrt{\kappa^2+1
+\frac{\alpha-1}{4(l+1+n)^2}}\ \right),\quad
\alpha>-3-2(k-|k|)k\end{gather} and
\beg\la{es12}\phi_{lm}^n=x\left(\frac{x^2+1}{x}\right)^{-n-l}
\exp\left(\frac{-E_n\arctan(x)}{2M}\right) P_n^{\left(-M-\frac{\ri
E_n}{4M},-M+\frac{\ri E_n}{4M}\right)}(\ri z)\end{gather} where
$z=\frac{1-x^2}{2x}$ and $M=n+l+1$.

Finally, for the Hamiltonian (\ref{Hef}) whose mass and potential
are fixed in  Item 9 of Table 1 we obtain the following energy
spectrum:
\beg\la{ev13}\begin{split}&E_n=4(l+1+n)^2\left(\kappa+\sqrt{\kappa^2-1
+\frac{\alpha+1}{4(l+1+n)^2}}\ \right),\\& \alpha>3-2(k+|k|)k,\
|\kappa|\geq1.\end{split}\end{gather}
 The corresponding eigenvectors are:
 \beg\la{es13}\phi_{lm}^n=x\left(\frac{x^2-1}x\right)^{-n-l}
 \left(\frac{x-1}{x+1}\right)^{\frac{-E_n}{2M}}P_n^{\left(-M-\frac{E_n}{4M},
 -M+\frac{E_n}{4M}\right)}(\tilde z)
 \end{gather}
 where $\tilde z=\frac{x^2+1}{2x}$.

Notice that potential and inverse mass terms fixed in the last lines of Tables 1 and 2 are singular at $x^2=\kappa+\sqrt{\kappa^2+1}$ and $x^2=\kappa+\sqrt{\kappa^2+1}$ correspondingly. However, solutions obtained in this section are regular, while the corresponding solutions $\psi=f^{-1}\Psi$ of the initial equations (\ref{se3}) are equal to zero in these points.

\section{Discussion}
The main goal of the present paper was to make the next step to the
complete classification of superintegrable PDM systems admitting
second order integrals of motion. Namely, we classify  rotationally
invariant systems having this property. The complete list of such
systems is presented in Tables 1 and 2. Thus  the first statement we
prove is that there are no other systems of the kind specified below
which are nonequivalent to the presented ones. The equivalence
relations of considered equations (\ref{se}), (\ref{H}) are given by
relations (\ref{et}), (\ref{IT}) and (\ref{et2}), or, more
generally, by arbitrary transformations belonging to the 3d
conformal group C(3) \cite{NZ}. Notice that in the latter case the
formal rotation invariance can be loosed.

Thus we present all nonequivalent rotationally invariant PDM systems
admitting second-order integrals of motion. The related integrals of
motion are presented explicitly in the fourth columns of the
mentioned tables.

In addition to its Hamiltonian, any of found  systems admits  four
algebraically independent integrals of motion, two of which commute
between themselves. The commuting integrals of motion are, say, $J_3$ and $J_1^2+J_2^2+J_3^2$ where $J_1, J_2$ and $J_3$ are components of angular momentum (\ref{J}). Two additional independent integrals of motion can be chosen as $Q_1$ and $Q_3$  or  as $Q_{33}$  and $Q_{12}$ for systems specified in Table 1 or Table 2 correspondingly. In other words, all these systems are maximally
superintegrable. Thus it is possible to formulate the second
statement: if the PDM system is rotationally invariant and admits at
least one second order integral of motion which is not a product of
its first order symmetries, it also admits two more such integrals
of motion and is maximally superintegrable.

Notice that almost all  rotationally invariant PDM systems admitting first order additional integrals of motion are maximally superintegrable too, see equations (\ref{fV2})--(\ref{fV4}) and (\ref{I2})--(\ref{I4}). The only exception is the system specified by equations (\ref{fV1}) which is superintegrable but not maximally superintegrable.

The next goal of this paper was to study the relations between the
superintegrability and supersymmetry of PDM systems. As expected
these relations appear to be very close. Namely, absolutely all
classified systems are also supersymmetric since their effective
potentials are shape invariant. The same is true for the first order systems (\ref{fV2})--(\ref{fV4}).

Many of the presented systems are characterized by analogous effective potentials. Indeed,   in the last columns of Tables 1 and 2 we can find eight cases of Eckart potential, six cases of Coulomb and the same number of oscillator potentials, etc. However, the systems with the same named effective potentials are essentially different. In some cases (like ones enumerated in items 1 and 2 of Table 2) the the same potentials correspond to different, i.e., direct and two step solution approaches and generate absolute different energy spectra. In the other cases the effective potentials have the same names but include different parameters. For example, comparing Item 2 of Table 1 and Item 1 of Table 2, in both cases we find the  Coulomb effective  potential appearing in the direct approach. However, in the case indicated in Table 1 the potential parameter $\mu$ is given by equation (\ref{mu}) while in the case presented in Table 1 we obtain equation (\ref{rey}) with $\mu=l$.

Some of the discussed systems have a rather
specific property which we call two fold shape invariance. Namely,
they possess extended hidden supersymmetry and  can be solved using
two different superpotentials. One  of such systems which is well known and is related to well known Coulomb-oscillator duality, is discussed  in
section 6.2. The other systems with the two fold shape invariance
are specified in Items 1--4 of Table 1  and Items 2--4  of
Table 2.

Thanks to their extended symmetries  the presented systems are
exactly solvable. Moreover, their supersymmetries make it possible
to construct solutions in a very easy way with using tools of SUSY
quantum mechanics. We find these solutions for the most complicated
systems whose hamiltonians are defined up to two arbitrary
parameters, see section 6.3.

We also present a simple algorithm for construction of exact
solutions for any of the considered systems, which reduces this
construction to a simple algebraic procedure since the related effective potentials are shape invariant, see section 6.1.  For alternative ways for
solving of one dimensional PDM Schr\"odinger equations see
\cite{tkach1}, \cite{tkach2}, \cite{jana}.

The rotationally invariant and superintegrable PDM systems were
discussed in numerous interesting papers by A. Ballesteros, A.
Enciso, F. J. Herranz, O. Ragnisco  and D. Riglioni. Selected papers
of this Spanish-Italian team are presented in the reference list,
see \cite{Bala2}--\cite{Rag3}.  Thanks to efforts of the mentioned
authors and their collaborators, such systems became a well studied
field. However, we believe that the present paper makes a
non-trivial contribution into this field in accordance with the
following arguments.

 In papers \cite{Bala2}--\cite{Rag3} the main accent is made
on classical Hamiltonian systems. Quantum mechanical systems are
considered also, but they appear as a result of quantizing of
classical ones.

The number of various second quantization procedures is rather extended. Moreover, starting with a particular classical
system, we can obtain a lot of its  quantum mechanical
counterparts, which in general are not equivalent between
themselves. In particular, the direct application of the so called
"PDM quantization" leads to loosing of the superintegrability
property of the systems considered in \cite{Rag2}. To keep this
property it is necessary to make a specific modification of
potentials \cite{Rag2}. Thus it is desirable to have a priori
classification of all non-equivalent superintegrable quantum
systems.

Just such classification for rotationally invariant PDM systems
admitting second order integrals of motion is given in the present
paper. Moreover, we write these systems in maximally simple forms,
which do not include arbitrary parameters whose values can be fixed
using equivalence transformations.  The list of superintegrable PDM systems given in the present
paper includes two new families of such systems including pairs of
arbitrary parameters, see the last items in both Tables 1 and 2.

 In the past the shape invariance of particular superintegrable
PDM systems was sporadically used   to construct their exact
solutions. We declare the existence of shape invariance {\it for all
superintegrable rotationally invariant systems admitting second
order integrals of motion} and fix the types of the corresponding
superpotentials. Moreover, we specify the position dependent masses
and potentials which correspond to Hamiltonians with
 two fold shape invariance.

In the present paper we restrict ourselves to $3d$ PDM systems.
However, our results admit a direct generalization to systems with
arbitrary dimension $d>3$.

It would be interesting to classify PDM Hamiltonians
 which lead to  other shape invariant
potentials, including potentials  with spin classified in \cite{N3}
and \cite{N4}. Some elements of such classification for systems with
constant masses can be found in \cite{NN1} and \cite{bdn}.

One more challenge is to extend the classification of the second
order integrals of motion to the case of generic PDM systems which
are not rotationally invariant. This work is in progress.

\renewcommand{\theequation}{A\arabic{equation}} %
\setcounter{equation}{0}
\appendix
\section{Solution of determining equations}
\subsection{Scalar integrals of motion}
Integrals of motion (\ref{Q}) are scalars w.r.t. rotations iff
$\mu^{ab}$ is reduced to tensor $\mu_1^{ab}$ given by equation
(\ref{mmu}). Substituting $\mu^{ab}=\mu^{ab}_1$ into equation
(\ref{mmmu1}) we obtain the following solution for $f$:
$$f=\varphi(x)$$
where  $\varphi(x)=\varphi_1(x)$ is an arbitrary function of $x$.

The corresponding integral of motion (\ref{Q}) takes the following
form:
\begin{gather}\la{s1}Q=-H-k_1{\bf J}^2+...\end{gather}
where the dots denote the term with the first and and zero order
differentials commuting with $H$. Since the first and second terms
in the r.h.s. evidently commute with $H$, the terms denoted by dots
also should be (first order) integrals of motion. Such integrals of
motion  had been classified in \cite{NZ} and presented above in
section 2.

\subsection{Vector integrals of motion}
Vector integrals of motion are associated with tensors $\mu^{ab}$
presented in (\ref{V1}) and (\ref{V2}). Moreover, versions
(\ref{V1}) and (\ref{V2}) correspond to vector and pseudovector
integrals of motion should be considered separately.

Let us start with tensor (\ref{V1}). Substituting it into
(\ref{mmmu1}) we come to the condition $f'=2xf$ and so $f=\alpha
x^2$. Using this expression for $f$ and substituting (\ref{V1}) into
(\ref{mmmu2}) we obtain the following equations for unknowns
$\xi^a$:
\begin{gather*}\xi^n_n=3\xi^cx_c, \qquad 3\left(\xi^a_b+\xi^b_a\right)-
2\delta^{ab}\xi^n_n=0\end{gather*}
and so $\xi^a=\tilde\lambda^ax^2-2x^a\tilde\lambda^bx^b$, where $\tilde
\lambda^a$ are arbitrary constants.

The remaining determining equations, i.e.,  (\ref{mmmu3}) and
(\ref{mmmu4}), generate the following conditions: $V=C$ and
$\tilde\lambda^a=\lambda^a$. In other words, we obtain known solutions
(\ref{fV1}) while the corresponding second order integrals of motion are
reduced to anticommutators of the first order symmetries (\ref{J}) and
(\ref{I1}). This result is trivial.

Consider now tensor (\ref{V2}). Substituting it into equations
(\ref{mmmu1}) and (\ref{mmmu2}) and equating coefficients for
linearly independent terms $\lambda^a, \ x^a\lambda_bx_b,\ x^ax^b,\
(x^a\lambda^b+x^b\lambda^a)$ and $\delta^{ab}\lambda_cx_c$   we come
to the following system of algebraic equations for differential
variables $f,\ \frac{f'}x$  and $\frac1x\left(\frac{f'}x\right)'$:
\begin{gather}\la{a1}\begin{split}&(\varphi+4\mu x^2)f-(\nu+\mu
x^2)xf'=0\\&(\varphi'-4\mu x)f-\left(\varphi-(\nu+\mu
x^2)\right)f'=0,\\&(6\varphi'+x\varphi''-16\mu x)f+4(1+\mu
x^2-\varphi)f'-\varphi
xf''=0,\\&2(\varphi''x-\varphi')f+2(\varphi-1-3\mu
x^2)f'+2(1-\varphi+\mu x^2)xf''=0,\\&2(2\mu x+\varphi')f+2\mu
x^2f'-(1+\mu x^2)xf''=0\end{split}\end{gather}

The compatibility condition for system (\ref{a1}) is given by
equation (\ref{phi}). Substituting the expressions (\ref{phi}) for
$\varphi$
 into one of equations (\ref{a1}) and going over all inequivalent
 versions of parameters $\mu$ and $\nu$ presented in (\ref{par1}) we
 obtain the following admissible pairs of functions $f$ and
 $\varphi$:
 \begin{gather}\la{a3}\begin{split}&f=x,\quad
 \varphi=-\frac1{4x^2}+\alpha,\quad \mu=0\\&f=x^3,\quad
 \varphi=1,\quad \mu=0\\& f=x^4,\quad \varphi=\mu=0,\\&f=\frac{x}{x\pm1}, \quad
 \varphi=\frac{1}{x\pm1},\quad \mu=0,
 \\&f=(1+\mu x^2)^2,\quad \varphi=0,\quad \mu=\pm1,\\
 &f=\frac{(x^2-1)^2x}{x^2-2\kappa x+\mu x^2},\quad \varphi=
 \frac{(1+\mu x^2)^2}{\nu-2\kappa x-\mu x^2},\quad \mu=\pm1.
 \end{split}\end{gather}

Thus we specify all functions $f$ for Hamiltonians
(\ref{H}) admitting vector integrals of motion. The generic form of
these integrals of motion is given by equations (\ref{Q}) and
(\ref{V2}) where
 \beg\la{eta}\eta=\lambda^ax_a\phi(x)\end{gather}
 and $\phi(x)$ is a yet unknown function.

 Substituting (\ref{Q}) and
(\ref{V2})  into the remaining determining equation (\ref{mmmu3})
and equating linearly independent terms proportional to $\lambda^a$
and $x^a\lambda^cx_c$ we come to the following system:
\begin{gather}\la{vde}\begin{split}&f(2x\phi+4\varphi'+x\varphi''-20\mu
x)+f'(x\varphi'-\varphi-10\mu x^2)+2(f''x-x^2V')(\mu
x^2-1)=0,\\&f(2\phi'x^2+\varphi'''x^2+4\varphi''x-4\varphi')+
f'(\varphi''x^2+14\mu x^2)-(f''x-f')(\varphi'x+6\varphi\\&-4+4\mu
x^2)-(f'''x^2-3f''x+3f')\varphi+2(\varphi-1+\mu
x^2)x^2V'=0.\end{split}\end{gather}

 Solving this system with all $f$ and $\varphi$ enumerated in
 (\ref{a3}) we obtain the corresponding potentials $V$ and functions
 $\phi$. Then, making transformation (\ref{last3}) we come to the
 results presented in Table 1.
 \subsection{Pseudotensor integrals of motion}
 Consider now determining equations (\ref{mmmu1})--(\ref{mmmu3}) with
 tensor $\mu^{ab}$ given by formula (\ref{a41}). They can be
 evaluated in complete analogue with the procedure outlined in the
 previous section, thus we will present the corresponding
 intervening results without comments.

 Equations
(\ref{mmmu1}) and (\ref{mmmu2}) result in the following system:
\begin{gather}\la{tt1}(\varphi'+8\mu x)f-
(2\mu x^2+\varphi)f'=0,\\\la{tt2}(2\varphi x-4\mu x^3)f+(\mu
x^4-1)f'=0,\\\la{tt3}(48\mu x+\varphi''x+8\varphi')f-(10\mu
x^2+5\varphi)f'-(1+\mu
x^4+x^2\varphi)\left(\frac{f'}{x}\right)'=0,\\\la{tt4}(2\mu
x+\varphi')f-4\mu x^2f'-(1-\mu
x^4)\left(\frac{f'}{x}\right)'=0,\\\la{tt5}(\varphi''x+\varphi')f+4\mu
x^2f'- x^2(\varphi+2\mu
x^2)\left(\frac{f'}{x}\right)'=0.\end{gather}

Equation (\ref{tt3}) is a linear combination of equations
(\ref{tt1}), (\ref{tt4}) and (\ref{tt5}) while (\ref{tt4}) and
(\ref{tt5}) are differential consequences of (\ref{tt2}) and
(\ref{tt1}) correspondingly. In other words, we can restrict
ourselves to the subsystem (\ref{tt1}) and (\ref{tt2}), whose
solutions are enumerated in the following formulae:
\beg\la{compc}\begin{split}&\varphi=\frac{2(\kappa
x^4-2x^2+\kappa)}{x^4-2\kappa x^2+\mu}, \quad
f=C_1\frac{(x^4-\mu)^2}{x^4-2\kappa x^2+\mu},\quad
\mu=\pm1\\&\varphi=-\frac{(x^2-\mu)^2}{x^2}\quad
f=C_2\frac{(x^4-\mu)^2}{x^2},\quad
\mu=\pm1\\&\varphi=-\frac1{x^2+k},\quad f=\frac{C_3}{x^2+k},\quad
\mu=0.\end{split}\end{gather} Here $C_1, C_2, C_3, \kappa$ and $k$
are integration constants and conditions (\ref{par1})--(\ref{par3})
are used.

Thus we fix all possible mass terms for Hamiltonians admitting
pseudotensor integrals of motion. The latter ones are given by
equations (\ref{Q}) and (\ref{a41}) with
\beg\la{eta1}\eta=\lambda^{ab}x_ax_b\phi(x).\end{gather}
Substituting (\ref{a41}), (\ref{compc}) and (\ref{eta1}) into the
last remaining determining equation (\ref{mmmu3}) and equating
linearly independent terms proportional to $\lambda^{ab}x_b$ and
$x^a\lambda^{cd}x_cx_d$ we obtain the following system of equations
for potentials $V$ and unknown function $\phi$:
\begin{gather}\la{last?}\begin{split}&(4\phi x-20\mu
x+12\varphi'+2x\varphi'')f+(2x\varphi'-20\mu x^2)f'\\&+2(\mu
x^4-1)\left(\left(\frac{f'}x\right)'-V'\right)=0,\\&f(2\phi'x+
6(x^3\varphi'''+\varphi''-x\varphi'))+ (40\mu x+\phi'+\phi'')f'\\&
-x(14\mu
x^2+7\varphi+x\varphi')\left(\frac{f'}x\right)'+2(x^2\varphi+2\mu
x^4)V'\\&-\frac1{x^3}(1+\mu
x^4+x^2\varphi)(x^2f''-3xf''+3f')=0.\end{split}\end{gather} Solving
this system with all $f$ and $\varphi$ presented in
 (\ref{a3}) we find the corresponding potentials $V$ and functions
 $\phi$. Then, making transformation (\ref{last3}) and rescaling
 independent variables  to simplify  expressions for $f$ and $V$  we come to the
 results presented in Table 2.

\subsection{Tensor integrals of motion}
 The last version of integrals of motion we should consider corresponds to tensor $\mu^{ab}$ of type (\ref{mue5}), i.e.,
  \begin{gather}\begin{split}&\mu^{ab}=\mu\left(\varepsilon^{acd}\lambda^{cb}x^d+ \varepsilon^{bcd}\lambda^{ca}x^d\right)\\&+\nu\left(2(x^a\varepsilon^{bcd} +x^b\varepsilon^{acd})\lambda^{dn}x^cx^n- (\varepsilon^{ack}\lambda^{bk}+ \varepsilon^{bck}\lambda^{ak})x^cx^2\right).\end{split}\la{a51}\end{gather}
 Substituting (\ref{a51}) into (\ref{mmmu1}) we come to the following equation:
 \begin{gather}4\nu xf=(\mu+\nu x^2)f, \quad \text{or}\quad  f=C(\mu+\nu x^2)^2.\la{a52}\end{gather}

 Using (\ref{a51}), (\ref{a52}) and equation (\ref{mmmu4}) we immediately find that $V=-6C\mu x^2$
 In other words, we recover Hamiltonians  (\ref{fV2}) and (\ref{fV4}) whilst the related symmetry operators (\ref{Q}) are products of the first order integrals of motion (\ref{I2}) and (\ref{I3}) correspondingly.

\end{document}